\documentclass[12pt]{article}
\usepackage{amsmath}
\usepackage{graphicx}
\usepackage{geometry}

\usepackage{natbib}

\usepackage[tight]{subfigure}
\ifx\pdfoutput\@undefined\usepackage[usenames,dvips]{color}
\else\usepackage[usenames,dvipsnames]{color}
\IfFileExists{pdfcolmk.sty}{\usepackage{pdfcolmk}}{} 
\fi
\usepackage[plainpages=false,pdfpagelabels,pagebackref=false,naturalnames=true,hyperindex=true,pdftitle={Self-organizing urban transportation systems},pdfauthor={Carlos Gershenson}]{hyperref}
\hypersetup{colorlinks=true,
urlcolor=Cerulean,linkcolor=BrickRed,citecolor=RoyalBlue,a4paper,
  pdfpagemode=None,
  pdfstartview=FitH}
\usepackage[all]{hypcap}

\begin{document}

\title{Self-organizing urban transportation systems}
\author{Carlos Gershenson$^{1,2}$ \\
$^{1}$ Instituto de Investigaciones en Matem\'aticas Aplicadas y en Sistemas \\
Universidad Nacional Aut\'onoma de M\'exico\\
Ciudad Universitaria\\
Apdo.\ Postal 20-726\\
01000 M\'exico D.F. M\'exico\\
%Tel. +52 55 56 22 36 19 \
%Fax +52 55 56 22 36 20 \\
\href{mailto:cgg@unam.mx}{cgg@unam.mx} \
\url{http://turing.iimas.unam.mx/~cgg} \\
$^{2}$ Centro de Ciencias de la Complejidad \\
Universidad Nacional Aut\'onoma de M\'exico\\
%$^{3}$Centrum Leo Apostel, Vrije Universiteit Brussel\\
%Krijgskundestraat 33 B-1160 Brussel, Belgium\\
%\href{mailto:cgershen@vub.ac.be}{cgershen@vub.ac.be} \ \url{http://homepages.vub.ac.be/~cgershen}
}
\maketitle

\begin{abstract}

Urban transportation is a complex phenomenon. Since many agents are constantly interacting in parallel, it is difficult to predict the future state of a transportation system. Because of this, optimization techniques tend to give obsolete solutions, as the problem changes before it can be optimized. An alternative lies in seeking adaptive solutions. This adaptation can be achieved with self-organization. In a self-organizing transportation system, the elements of the system follow local rules to achieve a global solution. Like this, when the problem changes the system can adapt by itself to the new configuration.

In this chapter, I will review recent, current, and future work on self-organizing transportation systems. Self-organizing traffic lights have proven to improve traffic flow considerably over traditional methods. In public transportation systems, simple rules are being explored to prevent the ``equal headway instability" phenomenon. The methods we have used can be also applied to other urban transportation systems and their generality is discussed.
\end{abstract}

\section{Introduction}

Traditional science, since the times of Galileo, Laplace, Newton, and Descartes, has assumed that the world is predictable \citep{Kauffman2008}. The implications of this assumption can be clearly seen with Laplace's demon: If an intellect had knowledge of the precise position and momentum of all atoms in the universe at a point in time, then it could use Newton's laws to describe all past and future events. This reasoning has been shown to be flawed for several reasons \citep{Binder:2008}. Among many of these, thermodynamics showed that there are irreversible processes where information is lost, so the demon would have no access to all past events. As for the future, deterministic chaos has shown that even when the ``laws" of a system can be known, this does not imply that the future state of a system can be predicted due to lack of precision. Moreover, complexity has also shown that the predictability of the world is a mistaken assumption, since \emph{interactions} between elements of a system generate novel information and constraints that make it impossible to know the state of a complex system beforehand.

This does not imply that we should abandon all hope of predictability. In urbanism, as in other areas, it is certainly desirable to have a certain foresight before designing and building a system. However, we need to accept that our predictability will be limited. Knowing this, we can expect that the unexpected will come. More practically, we can build systems that are able to \emph{adapt} to unforeseen situations while being \emph{robust} enough to resist them \citep{GershensonDCSOS}. One way of achieving this is exploiting the concept of self-organization \citep{GershensonHeylighen2003a}.

A system designed as self-organizing focusses on building the components of the system in such a way that these will perform the function or reach the goal of the system by their dynamic interactions. Like this, if the goal of the system changes, the components will be able to adapt to the new requirements by modifying their interactions. 

In the next two sections, work on self-organizing traffic lights and on the equal headway instability phenomenon in public transportation systems is exposed, respectively. In Section \ref{sec:future} future work is mentioned. A generalization of the use of self-organization in urban transportation systems closes the chapter.

\section{Self-organizing traffic lights}

Traffic lights in most cities are \emph{optimized} for particular expected traffic flows. In some cities, there are different expected flows for different hours, e.g. morning rush hour, afternoon rush hour, low traffic, etc.
The goal is to set green light periods and phases so that vehicles reach their destination with the least delay. In most cases, this approach is better than having no coordination between traffic lights \citep{Huang:2003}. However, it has several drawbacks. For example, most methods using the optimization approach assume averaged traffic flows, i.e. that there is the same probability to find a vehicle anywhere on a street. Still, vehicles in cities tend to aggregate in platoons (mainly because of red lights). Thus, there is a higher probability to find a vehicle close to another one, and there will be empty spaces between platoons. By neglecting this information, intersection capacity is wasted by giving green lights to empty streets or forcing some vehicles that are about to cross to stop and wait for the whole duration of a red light. Also, if streets happen to have a higher or lower density than expected for some reason (public event, road works, weather conditions, etc.), the traffic lights are ``blind" to a change in demand by some streets and force vehicles to wait unnecessarily.

\begin{figure}[htbp]
\begin{center}
\includegraphics[width=.45\textwidth]{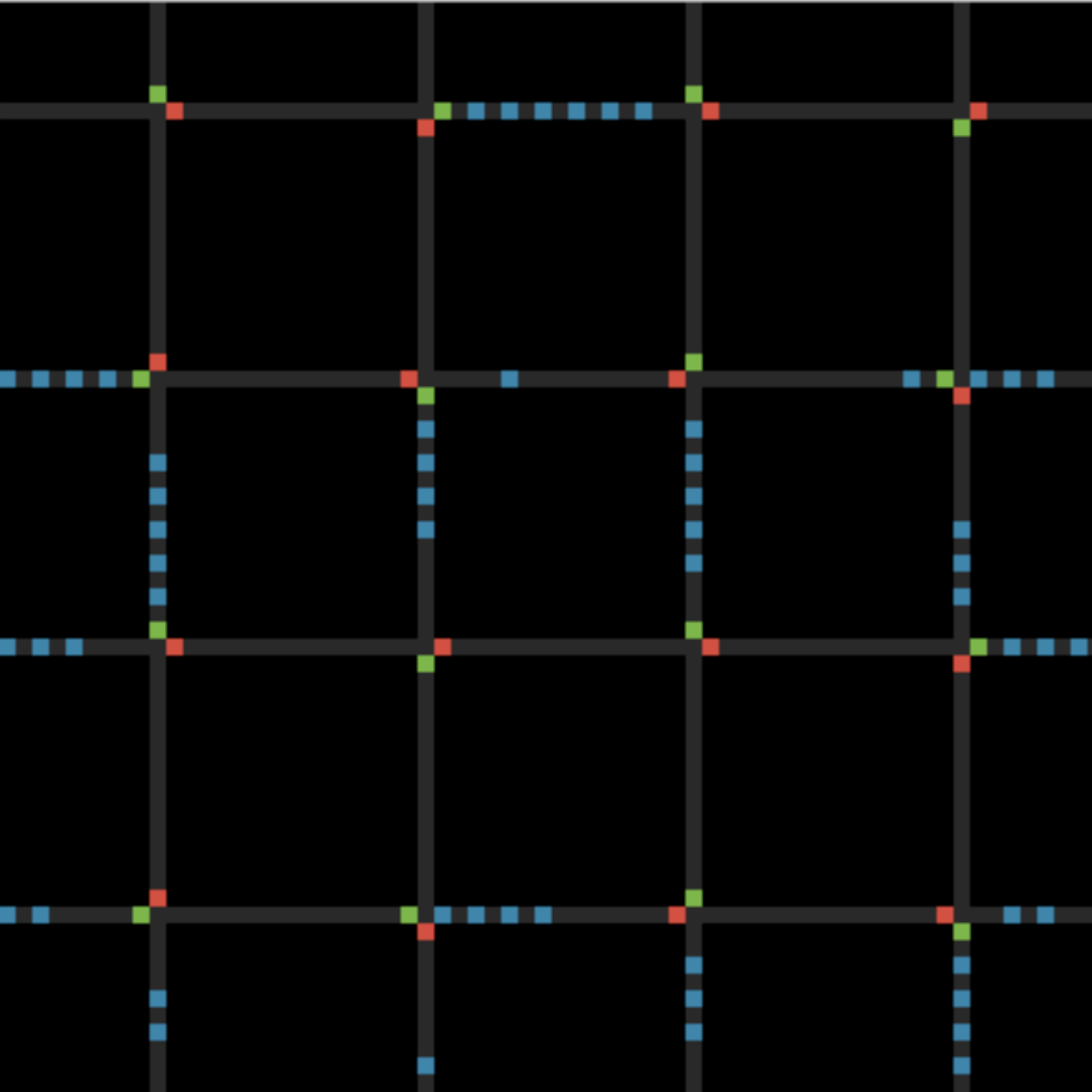}
\caption{Screenshot of the simulation. With the \emph{self-organizing method}, free flow is achieved in four directions for low densities ($\rho=0.1$ shown), as vehicles trigger green lights before reaching intersections.}
\label{fig:screenshot}
\end{center}
\end{figure}

An alternative lies in allowing traffic lights to self-organize to the current traffic situation, giving preference to streets with a higher demand \citep{Ball:2004,Gershenson2005,CoolsEtAl2007,GershensonRosenblueth2009}. The main idea is the following: each intersection counts how many vehicles are behind a red light (approaching or waiting). Each time interval, the vehicular count is added to a counter which represents the integral of vehicles over time. When this counter reaches a certain threshold, the red light switches to green. If there are few vehicles approaching, the counter will take longer to reach the threshold. This increases the probability that more vehicles will aggregate behind those already waiting, promoting the formation of platoons. The more vehicles there are, the faster they will get a green light. Like this, platoons of a certain size might not have to stop at intersections. There are other simple rules to ensure a smooth traffic flow, sketched in Table \ref{table:rules}. The method adapts to the current traffic density and responds to it efficiently: For low densities, almost no vehicle has to stop. For medium densities, intersections are used at their maximum capacity, i.e. there are always vehicles crossing intersections, there is no wasted time. For high densities, most vehicles are stopped, but gridlock is avoided with simple rules that coordinate the flow of ``free spaces" in the opposite direction of traffic, allowing vehicles to advance. The method reduces waiting times on average by 50\% \citep{CoolsEtAl2007} and prevents gridlocks at very high densities. 
For a detailed description of this \emph{self-organizing} method, please refer to \citet{GershensonRosenblueth2009}. The reader is invited to try a city traffic simulation available at 
\url{http://tinyurl.com/trafficCA}. A screenshot of a section of the simulation is shown in Figure \ref{fig:screenshot}.

\begin{table*}[htdp]
\caption{\emph{Self-organizing} traffic light rules. Inset: Schematic of an intersection, indicating distances $d$, $r$, and $e$ used for self-organizing lights.}
\label{table:rules}
\begin{center}
\vspace{0.2cm}
\fbox{
\parbox{.9\textwidth}{
%Box 1:

%\begin{figure*}
 \begin{center}
 %\label{fig:sensors}
   \includegraphics[height=30mm]{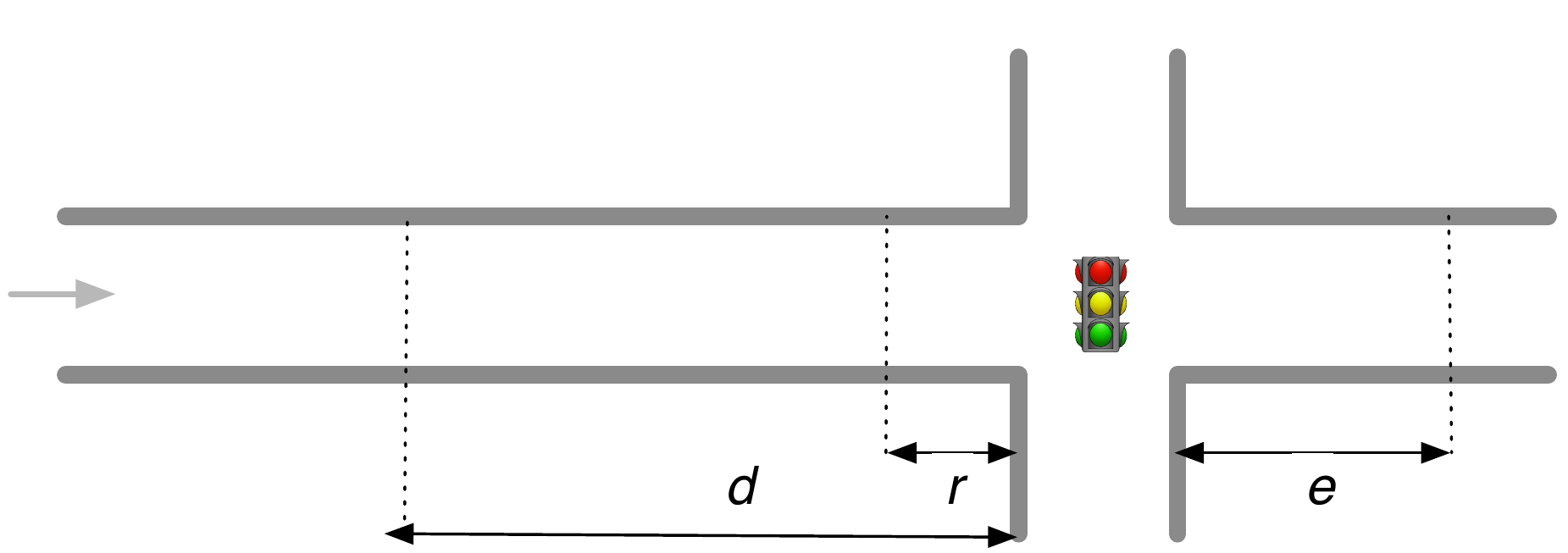}
 \end{center}
%  \caption{}
%\end{figure*}

{\small
\begin{enumerate}
\item On every tick, add to a counter the number of vehicles
approaching or waiting at a red light
within distance $d$.
When this counter exceeds a threshold $n$, switch the light.  (Whenever the light switches, reset the counter to zero.)
\item Lights must remain green for a minimum time $u$.
\item If a few vehicles ($m$ or fewer, but more than zero) are left to cross a green light at a short distance $r$, do not switch the light.
\item If no vehicle is approaching a green light within a distance $d$, and at least one vehicle is approaching
the red light within a distance $d$, then switch the light.
\item If there is a vehicle stopped on the road a short distance $e$
  beyond a green traffic light, then switch the light.
\item If there are vehicles stopped on both directions at a short distance $e$
  beyond the intersection, then switch both lights to red. Once one of the directions is free, restore the green light in that direction.
\end{enumerate}

}%small

}%parbox

}%fbox
%\vskip 12 pt

\end{center}

\end{table*}%

\begin{figure*}%[htp]
     \centering
     \subfigure{
          \label{fig:results_cityA}
          \includegraphics[width=.45\textwidth]{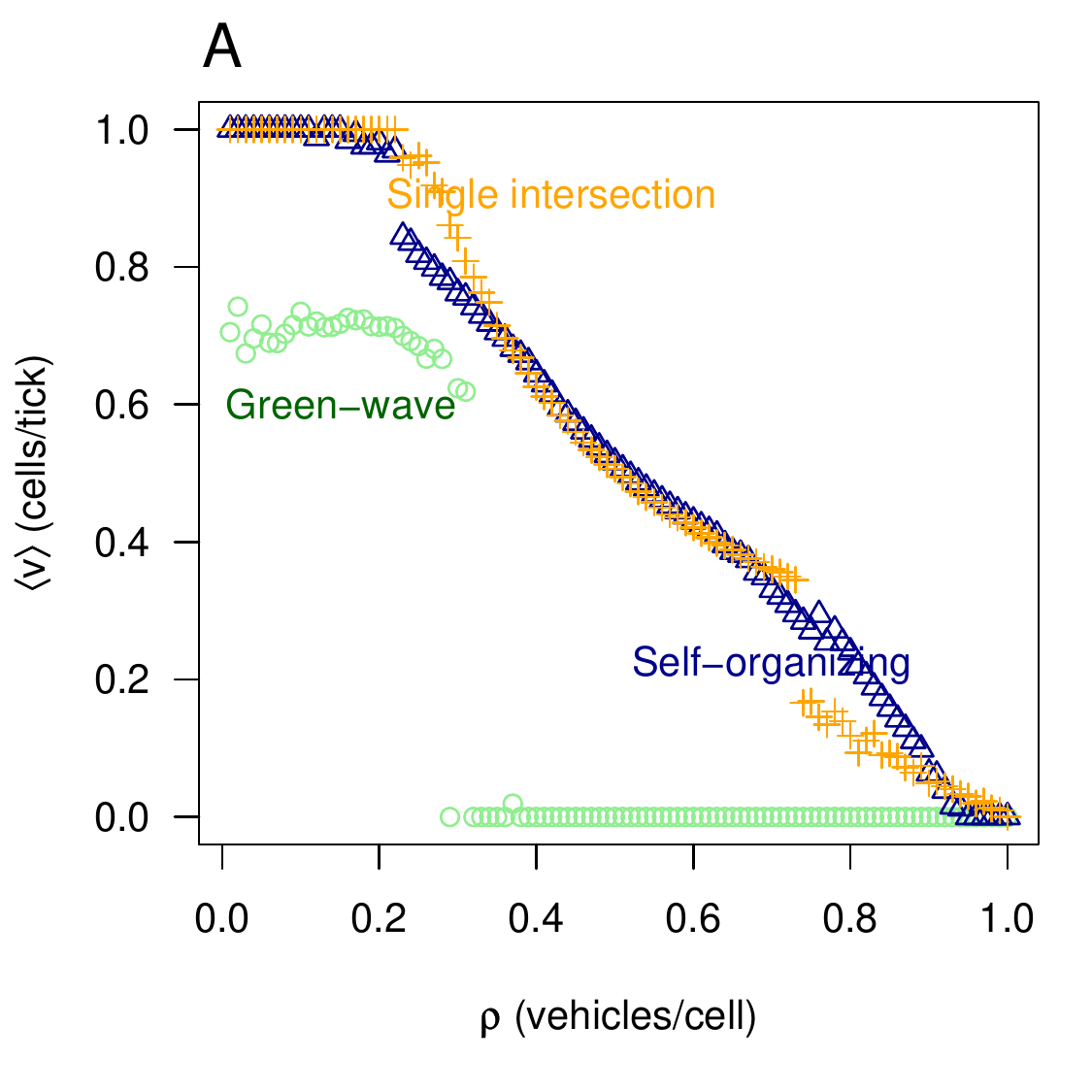}}
     \subfigure{
          \label{fig:results_cityB}
          \includegraphics[width=.45\textwidth]{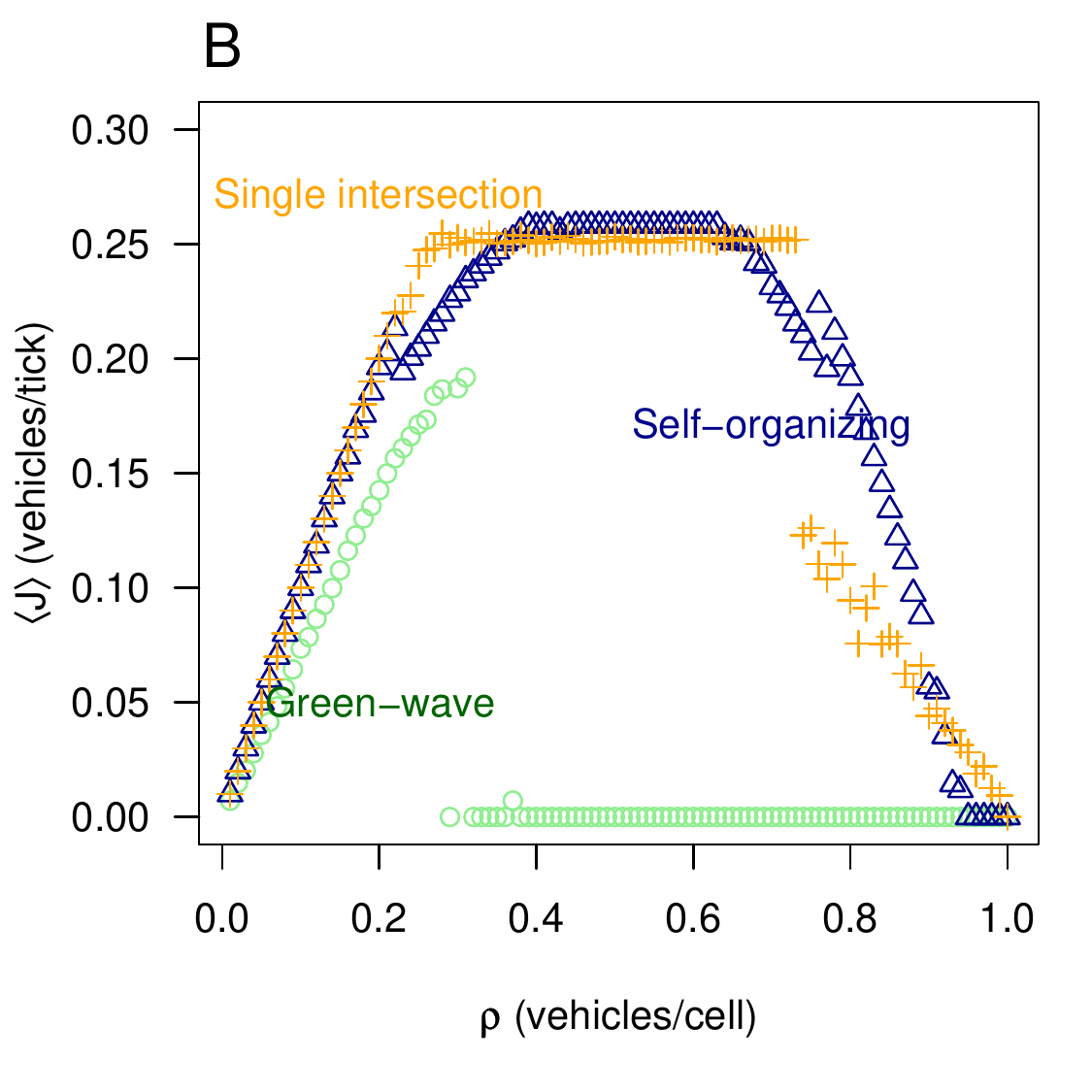}}
     \caption{Simulation results for a ten-by-ten city grid: (A) average velocity $\langle v\rangle$ and (B) average flux $\langle J\rangle$ for different densities $\rho$, \emph{green-wave} ({\Large $\circ$}) and self-organizing ($\bigtriangleup$) methods. For comparison, results for a single intersection ($+$) are also shown.}
     \label{fig:results_city}
\end{figure*}

The main results of our simulations of a simulated cyclic grid of ten by ten streets can be seen in Figure \ref{fig:results_city}. For comparison, we also show results for a single intersection to indicate the capacities of an intersection in our model depending on vehicle density $\rho$ ($\rho=0$ means empty streets, while $\rho=1$ means no free space). Since the problem to be solved is the coordination of traffic lights, the best possible solution would match the capacity of the single intersection. As it can be seen, the \emph{self-organizing} method matches or at least is very close to the maximum capacity for different densities\footnote{There are two particular cases where the \emph{self-organizing} method achieves better performance than a single intersection, which are artifacts of the simulation. For details, please refer to \citet{GershensonRosenblueth2009}}. The \emph{green-wave} method can support free-flow in two directions at low densities. However, the other two directions face uncorrelated traffic lights, having to stop every three blocks. Thus, the average velocity and flux is far from the case of the single intersection. Moreover, at medium densities queues block intersections upstream, leading to gridlocks.

The \emph{self-organizing} method achieves free-flow, i.e. $v=1$ for low densities. In other words, no vehicle stops. At medium densities, the maximum flux $J$ is reached, i.e. intersections are used at their maximum capacity, i.e. a vehicle is always crossing an intersection. Gridlock is prevented at high densities, unless intersections are blocked due to initial conditions.

The technology to implement the \emph{self-organizing} method is already available. However, further details should be dealt with before the method can be implemented, e.g. how to include pedestrians. The method also offers potential advantages for handling vehicles with priority (public transport, emergency, police, etc.). Weights can be added so that vehicles with priority can trigger by themselves green lights, i.e. behave as platoons, without interfering with the rest of the traffic.

\section{Public transportation systems and the equal headway instability phenomenon}

\begin{figure}[htbp]
\begin{center}
\includegraphics[width=15cm]{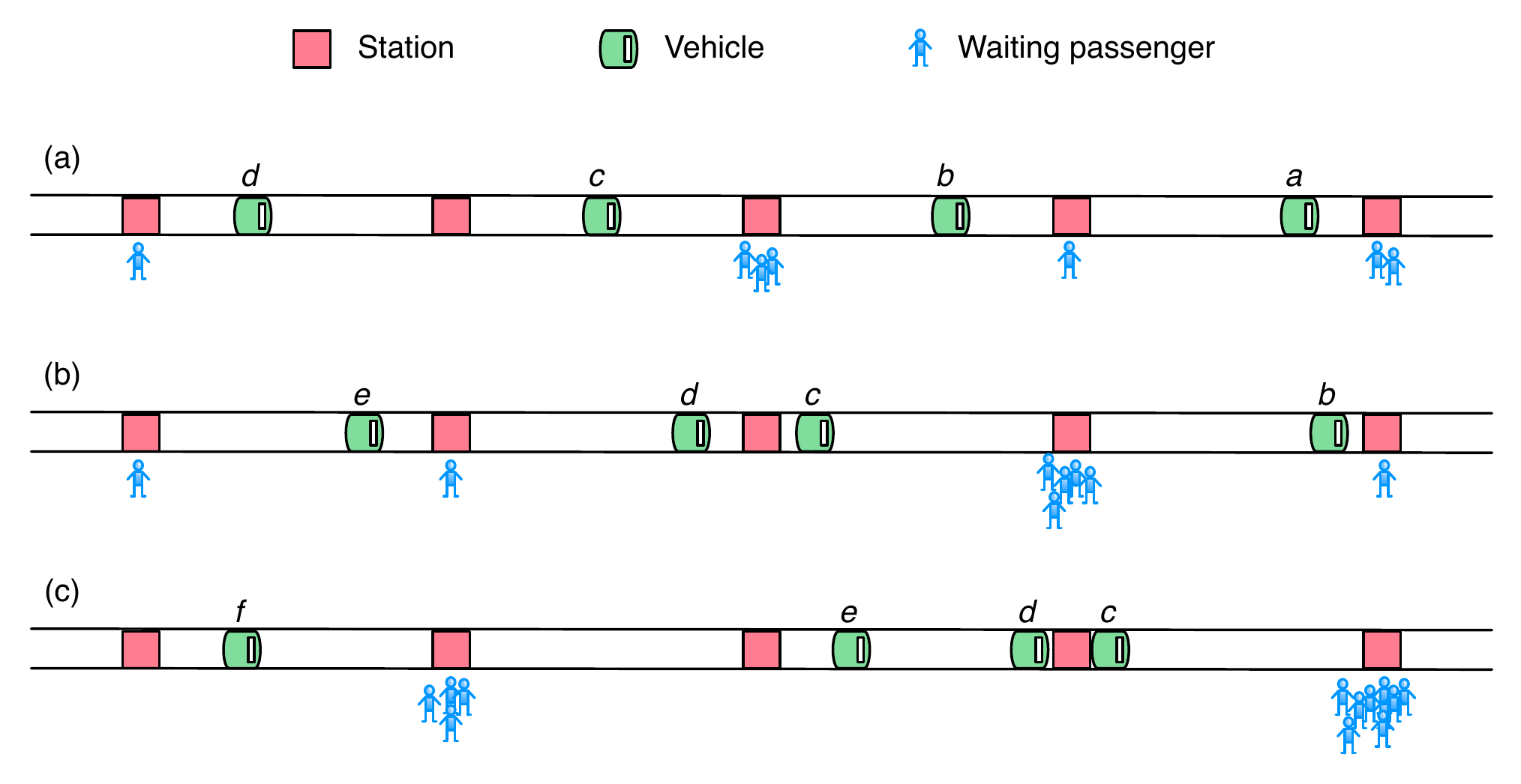}
\caption{Equal headway instability \citep{GershensonPineda2009}. a) Vehicles with a homogeneous temporal distribution, \emph{i.e.} equal headways. Passengers arriving at random cause some stations to have more demand than others. b) Vehicle $c$ is delayed after serving a busy station. This causes a longer waiting time at the next station, leading to a higher demand ahead of $c$. Also, vehicle $d$ faces less demand, approaching $c$. c) Vehicle $c$ is delayed even more and vehicles $d$ and $e$ aggregate behind it, forming a platoon. There is a separation between $e$ and $f$, making it likely that $f$ will encounter busy stations ahead of it. This configuration causes longer waiting times for passengers at stations, higher demands at each stop, and increased vehicle travel times. The average service frequency at stations is much slower for platoons than for vehicles with an equal headway.}
\label{HeadwayDeviation}
\end{center}
\end{figure}

Passengers in public transportation systems arriving randomly at stations are served best when the time intervals between vehicles---also known as the headway---is equal \cite[p. 133]{Welding:1957}. This is because equal headways imply regular intervals at stations between vehicles. Still, an equal headway configuration is not stable, as explained in Figure \ref{HeadwayDeviation}.

The equal headway instability phenomenon is present in most public transportation systems, including metros, trams, trains, bus rapid transit, buses, and elevators \citep{GershensonPineda2009}. 

\begin{figure}[htp]
     \centering
\includegraphics[width=0.6\textwidth]{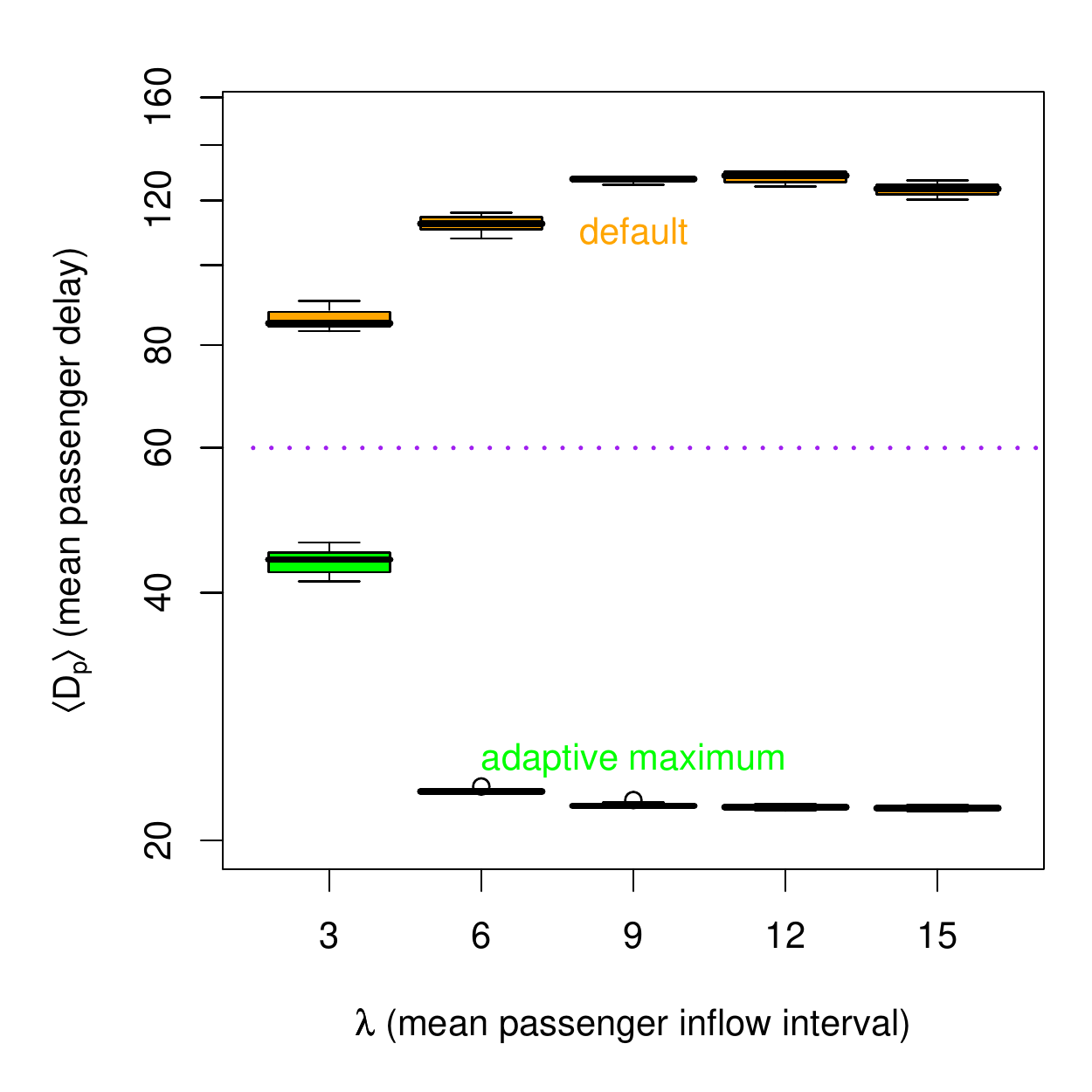}
          
       \caption{{ \textcolor{black}{Comparison of passenger delays $D_p$ between \emph{default} and \emph{adaptive maximum} methods, }}\textcolor{black}{varying mean passenger inflow intervals $\lambda$. For $\lambda>3$, the mean delays for the \emph{default} method are about six times larger than those for the \emph{adaptive maximum} method (notice logarithmic scale on $y$ axis).}}
     \label{fig:resultsVDefMax}
\end{figure}

We have developed a simple model of a metro-like system and implemented it in a multi-agent simulation (available at 
\url{http://tinyurl.com/EqHeIn} ). We tested different constraints to promote equal headways. However, different parameters were best for different densities. As an alternative, we implemented adaptive strategies, where the parameters are decided by the system itself depending on the passenger density. With this approach, equal headways are much more stable, improving the performance of the transportation system considerably \citep{GershensonPineda2009}. Figure \ref{fig:resultsVDefMax} shows results of our simulations in a scenario with five stations and eight vehicles. A \emph{default} method (with no constrains) always leads to equal headway instability. Even when there are several empty vehicles in the system, these do not improve the situation, since they aggregate behind a delayed one. An \emph{adaptive maximum} method constrains the times that vehicles spend at stations depending on passenger density and is able to maintain equal headways.

Even when these are encouraging results, the implementation of a technological solution is not enough, since transportation systems are \emph{used} by people. The \emph{social} aspect of the system cannot be neglected. In several systems, the main cause of the equal headway instability is passenger behavior. If appropriate measures are taken to promote ``positive" behaviors and avoid ``negative" behaviors, then the equal headway instability can be avoided and thus improve the efficiency of the transportation system.

For example, passengers should be discouraged from boarding crowded vehicles, since most probably they are leading a delayed platoon, with idling vehicles following behind. This could be encouraged with real-time information of the vehicle positions and/or expected time of arrival (one minute, ten minutes) and an indicator of their usage (empty, crowded, full).

Other behaviors that should be promoted are those that allow a fast boarding and exit of passengers. Examples of these are letting people exit before entering and not standing near doors during a trip. The vehicles and stations can also be designed to facilitate these behaviors, e.g. having dedicated doors for entering and exiting; an efficient distribution of doors, seats, bars, and other obstacles; monitors providing useful information for passengers, etc. Many of these suggestions seem obvious, but they are not followed by passengers in many cities. It can be more cost effective to promote efficient behaviors than implementing technological modifications to current systems.

\section{Future directions}
\label{sec:future}

The approach used to coordinate traffic lights and to promote equal headways can be applied in other systems. For example:

\begin{description}
\item[Coordination of public and private transport.] The combination of the studies presented above can be useful to improve the performance of bus rapid transit systems \citep{Levinson:2003}. These suffer from equal headway instability and are also affected by traffic lights. The implementation of self-organizing traffic lights with priority (treat one bus as a platoon) and measures to promote equal headways would agilize the flow of both buses and private vehicles. 
\item[Highway traffic.] Driving behavior can affect the capacity of highways. Different local rules and constraints can be explored to improve traffic flow.
\item[Crowd dynamics.] Pedestrian behavior in crowded and panic situations can lead to undesirable situations \citep{Helbing:2000}. Like with highway traffic, different changes in behavior can be explored to improve pedestrian flow and avoid accidents.
\end{description}

\section{Generalizing the use of self-organization}

The concept of self-orgnization can be used to design and build systems that are able to adapt to unforeseen situations in complex problem domains \citep{GershensonDCSOS}. The main idea is to build the components of a system in such a way that they will find by themselves the solution to a problem. Like this, when the problem changes, the system will be able to find a new solution.

Elements and systems can be described as agents with \emph{goals}. We can assign a value to a variable $\sigma \in [0,1]$ to represent the degree to which the goals of an agent have been met. The ``\emph{satisfaction}" of the agent is represented by $\sigma$.
 Agents in a complex system \emph{interact}. These interactions can have positive, neutral, or negative effect on the goals (and thus, $\sigma$) of other agents and of the system. If the interactions are negative, we can call them \emph{friction}. If they are positive, we can call them \emph{synergy} \citep{Haken1981}. If the friction between local interactions is minimized, then the satisfaction of the system will be maximized\footnote{Note that this is different from maximizing local satisfaction, which can lead to ``selfishness" or ``cheating" when a certain behavior can increase the local satisfaction in spite of reducing the system's satisfaction.} \citep[p. 41]{GershensonDCSOS}. \emph{Mediators} can be used to promote synergy and reduce friction. In this approach, the role of the designer lies in finding the appropriate mechanisms to steer agents in finding solutions at the system level, i.e. increase the system's satisfaction. Without defining a specific solution, the agents can adapt to changing problem domains.

This methodology, detailed in \citet{GershensonDCSOS}, was useful for developing the self-organizing traffic light controllers and the methods to promote equal headways. It can certainly be useful in other areas of urban transportation systems, as the ones mentioned in the previous section. For the case of the traffic lights, friction can be detected when cars have to stop. Thus, to avoid this friction, traffic lights need to ``get rid" as fast as possible of incoming vehicles (giving preference to streets with higher demand) and to prevent gridlocks (setting red lights to streets blocked ahead). Only with local interactions that reduce frictions, the satisfaction of the system is increased considerably compared with the \emph{green-wave} method. For the case of the public transportation systems, unstable headways cause friction, since they lead to some vehicles having excessive demand and others being idle. Thus, mediators that promote equal headways can be described as reducing friction that will lead to a better system performance, i.e. higher system satisfaction.

Since the interactions in a complex system generate novel information, equation-based approaches are not sufficient for these problem domains. Multi-agent simulations are a complementary alternative, because interactions are generated as simulations are run. Statistical results of such simulations can give insights on the functioning of complex systems. Simulations also allow the exploration and variation of different methods for reducing friction and promoting synergy that lead to better adaptive solutions. This has been done in our previous work and will be used in our future projects. 

Even when the focus here has been on urban transportation systems, the ideas presented could be applied in other areas of urbanism, e.g. adaptive urban planning and design.

%\section{Conclusions}

\section*{Acknowledgements}

I should like to thank Juval Portugali and his team for organizing the conference ``Complexity Theories of Cities have come of Age". Ideas on self-organization have been developed in collaboration with Francis Heylighen. Work on self-organizing traffic lights has been performed in collaboration with Seung Bae Cools, Bart D'Hooghe, Justin Werfel, Yaneer Bar-Yam, and David Rosenblueth. Work on the equal headway instability phenomenon has been made in collaboration with Luis A. Pineda.

\bibliographystyle{cgg}
\bibliography{carlos,sos,evolution,complex,traffic}

\end{document}